\documentstyle[12pt]{article}
\begin{document}
\voffset -2.5cm
\hoffset -1cm
\baselineskip 14pt
\title{Ground states of lattice gases with\\ ``almost'' convex repulsive
interactions}
\author{
Janusz J\c{e}drzejewski \\
Institute of Theoretical Physics,
University of Wroc{\l}aw
\thanks{Pl.\ Maksa Borna 9, 50--204 Wroc{\l}aw, Poland, e-mail:
jjed@ift.uni.wroc.pl}
and\\
Department of Theoretical Physics, 
University of {\L}\'{o}d\'{z}
\thanks{Ul.\ Pomorska 149/153, 90--236 {\L}\'{o}d\'{z}, Poland}\\
and \\
Jacek Mi\c{e}kisz\\
Institute of Applied Mathematics and Mechanics,\\
University of Warsaw
\thanks{Ul.\ Banacha 2, 02--097 Warszawa, Poland, e-mail:
miekisz@mimuw.edu.pl}
}
\date{}
\maketitle

\begin{abstract}

\noindent
To our best knowledge there is only one example of a lattice system
with long-range two-body interactions
whose ground states have been determined exactly:
the one-dimensional lattice gas with purely repulsive and strictly convex
interactions.
Its ground-state particle configurations do not depend on
the rate of decay of the interactions and are known as the
generalized Wigner lattices or the most homogenenous particle
configurations.
The question of stability of this beautiful and universal result
against certain perturbations of the repulsive and convex interactions
seems to be interesting by itself. Additional motivations for studying
such perturbations come from surface physics (adsorbtion on crystal
surfaces) and theories of correlated fermion systems (recent results on
ground-state particle configurations of the one-dimensional
spinless Falicov-Kimball
model).
As a first step we have studied a one-dimensional lattice gas whose
two-body
interactions are repulsive and strictly convex only from distance 2 on
while its value at distance 1 is fixed near its value at infinity.
We show that such a modification makes
the ground-state particle configurations sensitive to
the decay rate of the interactions:
if it is fast enough, then particles form
$2$-particle lattice-connected aggregates
that are distributed in the most homogeneous way.
Consequently, despite breaking of the convexity property,
the ground state exibits the feature known as the
complete devil's staircase.

\end{abstract}
most homogeneous configurations;
devil's staircase.

\newpage

\section{Introduction}

The present day equilibrium statistical mechanics is, without a great
exaggeration, a theory of lattice systems with translation-invariant
short-range interactions. The first step in a study of low-temperature 
properties of such systems amounts usually to determining
their ground-state configurations \cite{slawny}.
Provided the underlying interactions are short-range,
numerous methods of searching for ground-state configurations are
available.
In the case of one-dimensional lattice systems, there is even an
algorithmic method \cite{bundaru,miekisz_rad}.
In higher dimensions, let us mention the
powerful method of $m$-potential which leads to success in many
cases of interest \cite{slawny}.

However, if the translation-invariant interactions are
long-range ones, the situation is drastically different.
The exact results are scarce. To our best knowledge they are available
in two cases only:
(1) a version of one-dimensional Frenkel-Kontorova model \cite{aubry}
and
(2) a one-dimensional lattice-gas model \cite{hubb78,pokrovsky}.
This is in contrast with the case of continuous systems, where a number
of results by various authors are available; an overview of these
results and the corresponding references can be found in a review
paper by Radin \cite{radin}.

We would like to mention that even these one-dimensional lattice
cases are of great interest,
being no academic problems but important scientific issues,
well supported by the physics of quasi-one-dimensional materials
\cite{hubb79}, highly anisotropic layered systems \cite{fisher},
and adsorbtion of molecules on crystal surfaces
\cite{ishimura,sasaki,oleksy}.

The lattice-gas models with
purely repulsive and strictly convex two-body interactions emerged from
considerations of orderings of electrons in quasi-one-dimensional
conductors by Hubbard \cite{hubb78,hubb79} and orderings of
monolayers of atoms adsorbed on solids by Pokrovsky and Uimin
\cite{pokrovsky}. The model still appears to be a cameo in this field.
The periodic ground-state particle configurations of this model have
been characterized exactly in \cite{hubb78} and \cite{pokrovsky}.
Particles are distributed as far as possible from each other,
respecting restrictions imposed on their locations by the
underlying lattice. These configurations, called by Hubbard the
{\it generalized Wigner lattices} \cite{hubb78},
are independent of any further details of the interaction potential.
Moreover, the ground-state configurations exibit an interesting feature
known as the complete devil's staircase \cite{bak,aubry}

In view of such an impressing universality of the result obtained
independently by Hubbard and by Pokrovsky and Uimin, the question of
the stability of this result against some perturbations of the
interactions seems natural and interesting from the point of
view of statistical mechanics.
Moreover, the interest in this question is supported by other domains
of physics. In surface physics, a remarkable activity consists in
studying, both experimentally and theoretically, orderings of molecules
adsorbed on crystal surfaces. The phenomenon can be modelled by means
of one-dimensional lattice gases
\cite{ishimura,sasaki,oleksy} with specific two-body interactions,
which sometimes constitute certain local
perturbations of strictly convex repulsive interactions \cite{ishimura}.
Another motivation for studying effects of modifications of
repulsive and strictly convex two-body interactions stems from the recent
results concerning ground states of a version of the Falicov-Kimball
model \cite{falkim} -- the one-dimensional spinless
Falicov-Kimball model.
The model can be thought of
as a model of quantum itinerant electrons and classical localized
particles called f-electrons, nuclei or ions \cite{brandt,kennlieb},
with only on-site interactions
whose strength is the unique parameter of the model
-- a sort of the Ising model in the field of correlated fermion systems.
Such a system can be transformed into a classical lattice-gas
model with fairly complicated long-range and many-body interactions
\cite{kennlieb}.
Many years after the
discovery of the generalized Wigner lattices, Lemberger \cite{lemberger}
found the periodic ground-state configurations of the localized particles
in the large-coupling one-dimensional spinless Falicov-Kimball model.
He named them
the {\it most homogeneous configurations} because of the special role
they play in his ingenious procedure of
differentiation and integration of particle configurations.
It turns out that the configurations found by Lemberger are just the
generalized Wigner lattices.
Despite the complicated nature of
the effective interaction between electrons and ions, for some
details see \cite{grumacleb}, there is a numerical evidence
\cite{guj,gjl} (based on the approximate method of restricted phase
diagrams) that in the strong-coupling regime it can be
approximated by a repulsive and strictly convex two-body potential
that has the same set of periodic ground-state configurations.
Further studies of the ground-state
phase diagram of the one-dimensional spinless Falicov-Kimball model,
carried out in \cite{guj,gjl}, revealed for medium and small couplings
a number of
new families of good candidates for ground-state configurations.
Among them are the so-called $n$-{\em molecule most homogeneous}
configurations, found for the first time and studied in detail
in \cite{guj}.
Roughly speaking, they are obtained from the
most homogeneous configurations by replacing single particles by
lattice-connected aggregates of $n=2,3,\ldots$ particles.
The arguments in favor of the existence of
$n$-molecule most homogeneous configurations, provided in
\cite{guj}, are based not only on the approximate method of
restricted phase diagrams but also on the study of the interaction
energy between a few ions only. It has been found that for
large values of the unique parameter, for which the most homogeneous
configurations are the ground-state configurations,
the interaction energy between a few, say two or three, ions
is strictly convex.
On the other hand, for those values of the unique parameter for which
the $n$-molecule most homogeneous configurations are the
ground-state configurations,
the interaction energy between two or three ions
gets considerably lowered at short distances
(which apparently encourages forming $n$-molecules)
and consequently it becomes nonconvex at short distances.
These results led us naturally to the question whether it is possible
to obtain the $n$-molecule most homogeneous configurations as
ground-state configurations of a lattice gas with two-body
interactions, by modyfing, at short distances only, the two-body
interactions whose ground-state configurations are the most homogeneous
ones.

The answer to this question is in the affirmative, at least in the
case $n=2$ which we have considered as a first step of our
investigations. Namely, we prove that repulsive interactions whose value
at distance 1 is set near their value at infinity, that are strictly convex
from distance 2 on, and whose rate of decay is fast enough
(according to a simple criterion), do the job.

Needless to say that we believe the sort of result we
obtained to be valid for any $n$, however a proof along the lines of
the case $n=2$ seems to be technically complicated.

The paper is organized as follows. In Section 2, we introduce the system
under consideration and give basic definitions. 
Then, in Section 3 we formulate our hypothesis, provide examples
that reveal problems that have to be solved on the way towards the final
result, and draw our strategy of proving the hypothesis.
A relatively simple part of this strategy, Lemma 0, is proved
in this section.
After that we prove our main theorem, Theorem 1,
by means of Lemma 0, 1, and 2.
The proofs of Lemma 1 and 2 are given in Section 4
and constitute the major part of the paper.
Finally, in Section 5 we provide a discussion of the obtained results,
limitations and possible extensions.
Some technical definitions and statements that are used
throughout the text are collected in the Appendix.

\section{Lattice-gas models and devil's staircases}

A classical lattice-gas model, considered here,
is a system in which every site of a one-dimensional lattice ${\bf Z}$
can be occupied by one particle or be empty.
Then, an infinite-lattice configuration is an assignment of
particles to lattice sites, i.e., an element of 
$\Omega = \{1,0\}^{{\bf Z}}$. 
If $X \in \Omega$ and $\Lambda \subset {\bf Z}$,
then we denote by $X_{\Lambda}$ a restriction of $X$ to $\Lambda$.
We assume that the particles interact only through two-body forces
and to a pair of particles at lattice sites $i$ and $j$,
whose distance is $|i - j|$, we assign the translation-invariant
{\em interaction energy} $V(|i - j|)$. The corresponding two-body
{\em potential} reads: $V(|i-j)|)s_{i}(X)s_{j}(X)$,
where $\{s_{i}(X), i \in {\bf Z} \}$ are
occupations of sites in a configuration $X$;
$s_{i}(X)$ assumes value $1$ if in the configuration $X$
the site $i$ is occupied by a particle and otherwise value $0$.
In terms of the above defined potential, the Hamiltonian of our system
in a bounded region $\Lambda$ amounts to the sum of the potential
over all pairs of sites having nonvoid intersection with $\Lambda$:
\begin{equation}
\label{H}
H_{\Lambda}(X)=
\sum_{ \{\{i,j\}: \{i,j\} \cap \Lambda \neq \emptyset \} }
V(|i - j|)s_{i}(X)s_{j}(X).
\end{equation}

For $Y,X \in \Omega$, we say that $Y$ is a {\it local excitation} of $X$,
and write $Y \sim X$, if there exists a bounded $\Lambda \subset {\bf Z}$
such that $X = Y$ outside $\Lambda.$ 

For $Y \sim X$, the {\it relative Hamiltonian} is defined by
\begin{equation}
\label{rel_H}
H(Y,X)=\sum_{ \{\{i,j\}: \{i,j\} \cap \Lambda \neq \emptyset \} }
\left( V(|i - j|)s_{i}(Y)s_{j}(Y) - V(|i - j|)s_{i}(X)s_{j}(X) \right).
\end{equation}

We say that $X \in \Omega$ is a {\it ground-state configuration} of $H$
if $$H(Y,X) \geq 0 \; \; for \; \; any \; \; Y \sim X.$$

That is, we cannot lower the energy of a ground-state
configuration by changing it locally.

The {\it energy density} $e(X)$ of a configuration $X$ is 
\begin{equation}
\label{e}
e(X)=\liminf_{\Lambda \rightarrow {\bf Z}}
\frac{H_{\Lambda}(X)}{|\Lambda|},
\end{equation}
where $|\Lambda|$ is the number of lattice sites in $\Lambda$.
In a similar way we define the {\it particle density} in a
configuration $X$, denoted $\rho (X)$,
simply the Hamiltonian in (\ref{e}) has to be
replaced by the number of particles in the region $\Lambda$.

It can be shown that any  ground-state configuration has the minimal
energy density which means that local
conditions present in the definition of a ground-state
configuration enforce global minimum of the energy density
\cite{sinai}.

For certain values of external parameters, like the chemical potential
or the particle density (depending on the used Gibbs ensemble),
our models do not have periodic ground-state configurations.
However, for any fixed value of such an external parameter,
all ground-state configurations belong to one
{\em local isomorphism class}. It means that 
they cannot be locally distinguished one from another. Every local 
pattern of particles present in one 
ground-state configuration appears in any other within a bounded distance.
More formally, there exists a unique {\em translation-invariant
probability measure} supported by ground-state configurations.
It is then necessarily the zero-temperature limit of
{\em equilibrium states}
(i.e., translation-invariant Gibbs states) \cite{enter}.

In this paper, by the {\em ground state} of a model we mean
precisely the above defined probability measure.

If a system has a unique periodic ground-state 
configuration and its translations (this happens in our models
for those values of the chemical potential that fix a rational
particle density),
then a unique ground-state measure assigns an
equal probability to all these translations. For example, the
Ising antiferromagnet has two alternating ground-state
configurations but only one ground-state measure which assigns
probability $1/2$ to both of them.   

In the nonperiodic case, a probability ground-state measure $P$ gives
equal weights to all ground-state configurations and can be obtained as
a limit of averaging over a given ground-state configuration $X$
and its translations $\tau_{{\bf a}}X$ by  lattice vectors ${\bf
a} \in {\bf Z}$: $P= \lim_{\Lambda \rightarrow {\bf Z}}
\frac{1}{|\Lambda|} \sum _{{\bf a} \in \Lambda} \delta
(\tau_{{\bf a}} X)$, where $\delta (\tau_{{\bf a}}X)$ is a
probability measure assigning probability $1$ to $\tau_{{\bf a}}X$.

One more remark concerning ground states discussed here is in order.
We consider exclusively the ground-state measures that are 
{\em strictly ergodic} \cite{kakutani,keane}. In particular,
every ground-state configuration in their support has uniformly
defined densities of all local particle configurations.
Moreover, if a local particle configuration occurs, then it occurs
with a positive density.
That is to say, we do not consider ground-state configurations
with interfaces (like kink ground-state configurations in the Ising model).

We say that a set $\Lambda \subset {\bf Z}$ is {\it lattice-connected}
if for every pair of lattice sites $i,j \in \Lambda$  there is a sequence
$i_{1}, \ldots, i_{n}$ such that $i_{1}=i$, $i_{n}=j$ and 
$i_{k}$, $i_{k+1}$, $k=1,\ldots, n-1$ are the nearest-neighbor sites.

In the sequel, in order to describe some local configurations
$X_{\Lambda}$, $Y_{\Lambda}$, for a lattice-connected $\Lambda$,
we find it convenient to introduce on ${\bf Z}$
a {\it coordinate-axis} with lattice sites located at integer points
and to call the positive direction the right one while
the opposite direction -- the left one.
The {\it environment of $\Lambda$}, ${\bf Z} - \Lambda$,
splits naturally into the {\it left environment}, consisting
of the sites preceding $\Lambda$, and the {\it right environment},
consisting of the sites following $\Lambda$. It will also be
convenient to set the zero of the axis at the last occupied site of
the left environment so that the position of a particle in $\Lambda$ is
positive and coincides with its distance to the left environment.
The coordinate-axis introduced will be briefly called $x$-{\em axis}.

We assume that the {\it interaction energy} of two particles separated by
distance $r$, $V(r)$, is summable and for 
$r \geq r_{0}$
it is positive and strictly convex (see the Appendix), hence decreasing.
Such an interaction energy is denoted by $V_{r_{0}}(r)$.
It is convenient to normalize $V_{r_{0}}(r)$ in such a way that
$\lim_{r \rightarrow \infty} V_{r_{0}}(r) = 0$.

Then, we define the interaction energy of a particle at a site $r$
in $\Lambda$
with the left environment of $\Lambda$, $V_{r_{0}}^{L}(r)$,
as the sum of interaction energies $V_{r_{0}}(r - j)$ over all
occupied sites $j$ in the left environment.
The properties of the interaction
energy $V_{r_{0}}$ imply that, for $r \geq r_{0}$, the function
$V_{r_{0}}^{L}(r)$ is positive, strictly convex, and decreasing.

In what follows some particle configurations play a distinguished part.
Among local configurations that appear frequently in our considerations
are {\em atoms}, i.e., occupied sites whose nearest-neighbor sites are 
empty. The {\em location of an atom} is identified with the location of
the occupied site. Another local configuration is a lattice-connected set
of $n$, $n \geq 2$, occupied sites whose left and right nearest-neighbor 
sites are empty. It is called the $n$-{\em molecule}.
The {\em location of a $n$-molecule} is identified with the location of
the first particle of the molecule, i.e., the one with the smallest
$x$-coordinate. Consequently, the
{\em distance between two $n$-molecules}
is identified with the distance between the first
particle of one $n$-molecule and the first
particle of the other $n$-molecule.

However, our attention is focused on global particle configurations 
which can be characterized as follows.
For every particle density $\rho$,
there is a unique sequence of natural numbers $d_{n}$, such that the 
separations between any pair of $n$-th nearest-neighbor particles are
$d_{n}$ or $d_{n}+1$. Such configurations
have been called the {\em generalized Wigner lattices} \cite{hubb78} or the
{\em most homogeneous configurations} \cite{lemberger}.
If the particle density is rational, then the corresponding most
homogeneous configurations are periodic
(the particle locations can be given by means
of a construction given for instance in \cite{hubb78}) while for irrational
densities they are nonperiodic.

At least in this paper, the main reason of interest in the most
homogeneous  configurations stems from the following theorem:\\[2mm]
\noindent
{\bf Theorem 0} (Most homogeneous ground-state configurations)\\
In the canonical ensemble,
i.e., for a given particle density $\rho$,
the ground-state configurations of a lattice gas (\ref{H})
with an interaction energy $V_{1}$
are the most homogeneous configurations.\\

This statement has been proven (or at least a proof has been outlined)
by Hubbard \cite{hubb78} and Pokrovsky and Uimin \cite{pokrovsky}.
The ground-state phase diagram in the grand-canonical ensemble has been
calculated heuristically by Bak amd Bruinsma \cite{bak}
while a proof for the related Frenkel-Kontorova model
has been provided by Aubry \cite{aubry} and adapted to the lattice-gas
model case by Mi\c{e}kisz and Radin \cite{mie1}; it is outlined below.

In the grand-canonical ensemble, to find the energy density
of a ground state we have to minimize
\begin{equation}
\label{free_en}
f(\rho)=e(\rho)- \mu \rho.
\end{equation}  
Now, $e(\rho)$, i.e., the ground-state energy density for the particle
density $\rho$, is differentiable at every irrational $\rho$ and is
nondifferentiable at any rational $\rho$ \cite{aubry,mie1}. However,
as a convex function,
it has a left derivative $d^{-}e(\rho)/d\rho$ and a right derivative 
$d^{+}e(\rho)/d\rho$ at every $\rho$. It follows, that to have a ground
state with an irrational density $\rho$ of particles, one has to fix
$\mu(\rho)=de(\rho)/d\rho$. For any rational $\rho$,
there is a closed interval of chemical potentials
$[d^{-}e(\rho)/d\rho, d^{+}e(\rho)/d\rho]$, where the most homogeneous
configurations of density $\rho$ are the ground-state configurations.
One can show that the sum of lengths of these intervals amounts to the
length of the interval begining at the end of the half-line, where the
vacuum is the only ground-state configuration, and ending at the
begining of the half-line, where the completely filled configuration
is the only ground-state configuration.

As we have already mentioned, for any rational $\rho$, there is a unique
(up to translations) periodic ground-state configuration with that density
of particles - there is a unique ground-state measure. For any irrational
$\rho$, there are uncountably many ground-state configurations which are
the most homogeneous configurations. It has been shown in \cite{mie2}
that there is still the unique ground-state measure supported by them.

The particle density versus the chemical potential of particles,
$\rho(\mu)$,
is constant in each set of this partition. Moreover, it is a continuous
function on the real line and is inversion symmetric with respect
to the point $(\mu_{0}, \rho (\mu_{0}))$, where $\mu_{0}$ is the
chemical potential for which the free energy density is hole-particle
symmetric ($\rho (\mu_{0})=1/2$). The curve
$\rho(\mu)$ is classified as a fractal one and named the {\em complete
devil's staircase} \cite{aubry}.

The last remark is that without any loss of generality we can restrict
our considerations to systems whose particle density $\rho$ does not exceed
$1/2$. Then the most homogeneous configurations consist exclusively
of atoms.

\section{Basic ideas and the main result}

Whether we follow the argument of Hubbard \cite{hubb78}
(based on a version of our Lemma A1) or the
argument of Pokrovsky and Uimin \cite{pokrovsky} that proves
Theorem 0, we find that such a proof consists essentially of two
stages. In the first stage we ``chop''
configurations that contain $n$-molecules with some
$n=2,3,\ldots$ off the set of all configurations with particle density
$\rho$,
so we are left only with configurations that are
composed of atoms whose density is $\rho$.
The second stage is like a ``fine tuning'' that amounts to precise
adjusting the positions of atoms in order to minimize the energy
density.

Is the strategy outlined above useful if the two-body strictly convex
for $r \geq 1$ interaction energy $V_{1}$, that appears in Theorem 0,
is replaced by an interaction energy $V_{r_{0}}$
which is strictly convex only for $r \geq r_{0}$, $r_{0}=2,3,\ldots$?

Consider first the second stage. Suppose that the set of
configurations composed of atoms only is replaced by the set of
configurations that consist exclusively of $n$-molecules with fixed
$n$, separated by at least distance $r_{0}+n-1$. Let the particle density
of these $n$-molecule configurations be $\rho$. This class of
configurations we denote by $C^{n}_{r_{0},\rho}$.
Clearly, the problem of determining the ground-state configurations
in $C^{n}_{r_{0},\rho }$ can be reduced to the analogous problem
but in the class $C^{1}_{r_{0},\rho /n}$ and with $V_{r_{0}}$
replaced by the {\em effective two-body interaction between
$n$-molecules}, $V^{(2)}_{r_{0}}$.
This effective interaction can be naturally defined
as the sum of interactions $V_{r_{0}}$ between
ordered pairs of particles such that the first member of a pair
belongs to one $n$-molecule while the second member -- to the other
$n$-molecule:
\begin{eqnarray}
\label{V2}
V^{(2)}_{r_{0}}(r) = nV_{r_{0}}(r) +
(n-1) \left( V_{r_{0}}(r+1) + V_{r_{0}}(r-1) \right) + \nonumber \\
(n-2) \left( V_{r_{0}}(r+2) + V_{r_{0}}(r-2) \right) + \ldots
+ \left( V_{r_{0}}(r-n+1) + V_{r_{0}}(r+n-1) \right),
\end{eqnarray}
where $r$ stands for the distance between the two considered $n$-molecules.
Since the function $V^{(2)}_{r_{0}}(r)$ is a sum of functions
that are strictly convex for $r \geq r_{0}$,
it is also strictly convex for $r \geq r_{0}$.
This implies that in the ground-state configurations, the first
particles of $n$-molecules form the most homogeneous configurations
with particle density $\rho /n$, and the same holds true for the second
particles of $n$-molecules, etc.
Thus, it is natural to call the obtained ground-state configurations
the {\em n-molecule most homogeneous configurations} of particle
density $\rho$.
We summarize the above considerations in the following lemma:\\[1 mm]

\noindent {\bf Lemma 0}
(Ground-state configurations restricted to $C^{n}_{r_{0},\rho }$)\\
In a lattice gas (\ref{H}) with an interaction energy $V_{r_{0}}$,
the ground-state configurations restricted to $C^{n}_{r_{0},\rho }$
are the $n$-molecule most homogeneous configurations.\\[1 mm]

Having generalized succesfully the stage two, it is tempting to turn to
the stage one. Can we modify $V_{1}$ in such a way that the
ground-state configurations are in $C^{n}_{r_{0},\rho}$?
The general suggestion that comes from studies of the
Falicov-Kimball model \cite{guj} is to set the values of
the two-body interaction energy at short distances close to zero.
One might expect that setting $V_{1}(1)=0$ will force the system to
form 2-molecules exclusively in the ground-state configurations.
However, in order to arrive at such a result, one has to deal with
other features of the interaction energy.
At this point it is instructive to turn to examples.
\vspace{2mm}

\noindent
{\bf Example 1}\\
Consider the convex two-body interaction energy $V$ that in some units
is given by $V(1)=0$, $V(2)=4$, $V(3)=2$, $V(4)=1$, $V(r)=0$,
for $r \geq 5$,
and two periodic configurations (period 17) of particle density
$\rho=8/17$ whose elementary cells are of the form:
$[\bullet \bullet \circ \circ \circ \bullet \bullet\bullet
\circ \circ \circ  \bullet \bullet\bullet \circ \circ \circ ]_{1}$
and
$[\bullet \bullet \circ \circ \bullet \bullet \circ \circ
\bullet \bullet \circ \circ \bullet \bullet \circ \circ \circ ]_{2}$,
where $\bullet$ stands for a particle while $\circ$ -- for an empty
site.
The energy per elementary cell in the first case is
$2 V(2)$ + $3 V(4) = 11$, while in the second case it is
$3 V(3)$ + $7 V(4) = 13$. Thus the configuration $[.]_{2}$,
which consists exclusively of 2-molecules, looses against the
configuration $[.]_{1}$, which consists of 2-molecules and 3-molecules.
While the interaction energy chosen is not strictly convex,
it is easy to see that the above result remains true if we do not set
$V(r)=0$ for $r \geq 5$ but extend $V(r)$ to some $V_{1}$, with
$V_{1}(r)$ positive and arbitrarily small from $r=6$ on.
We deal with the above problem in Lemma 2 which tells us that
to exclude $n$-molecules with $n\geq 3$ from competition one should
impose a condition on the relative strength of $V_{1}(2)$ with respect
to $V_{1}(r)$ with $r\geq 3$. Namely, it is sufficient to require that
the energy of the ``tail'' of $V_{2}$, defined as
$W=\sum_{r=3}^{\infty} V_{2}(r)$, is weak enough compared to
$V_{2}(2) + V_{2}(1)/2$, i.e., $V_{2}(2) + V_{2}(1)/2 \geq 7W/2$.
\vspace{2mm}

\noindent
{\bf Example 2}\\
This example shows that configurations containing
atoms are unlikely to be the ground-state ones.
Let the two periodic configurations of particle density $\rho = 3/11$
be of the form:
$[\bullet \bullet (5 \circ) \bullet \bullet (5 \circ)
\bullet \bullet (6 \circ)]_{3}$ (period 22) and
$[\bullet \bullet (4 \circ) \bullet (4 \circ)]_{4}$ (period 11).
The notation ``$(5 \circ)$'' stands for five empty sites in a row, etc.
If $V(r)=0$ for $r>8$, then the energy per cell of 22 sites amounts
to $2 V(6)$ + $5 V(7)$ + $4 V(8)$, in the case of $[.]_{3}$,
and to $4 V(5)$ + $4 V(6)$, in the case of $[.]_{4}$.
Now, let the interaction energy $V_{2}$ be chosen to coincide with
the finite-range interaction energy defined as follows:
$V(1)$ is nonpositive, $V(2)=28$, $V(3)=21$, $V(4)=15$,
$V(5)=10$, $V(6)=6$, $V(7)=3$, $V(8)=1$, in some units,
while for $r \geq 9$, $V_{2}(r)$ is an arbitrarily small positive extension
of this function to a strictly convex function.
For such $V_{2}$ the configuration $[.]_{3}$ wins overwhelmingly.

Summing up, in order to implement the stage two in the case of a
nonconvex interaction energy $V_{2}$, we have to force the system,
by modyfying its interactions,
to make the configurations containing $n$-molecules with
$n \geq 3,4,\ldots$ unfavourable energetically,
and to show generally that for such interactions the
configurations containing atoms cannot be among the
ground-state ones.
The latter problem is dealt with in Lemma 1.

Lemma 1 applied for instance to the configuration $[.]_{4}$
tells us that,
for any $V_{2}$ such that $V_{2}(1) \leq 0$,
if we take twice
as large elemetary cell and rearrange the two atoms and the
2-molecule separating them, so that they form
two 2-molecules distributed as follows:
$[\bullet \bullet (6 \circ) \bullet \bullet (4 \circ)
\bullet \bullet (6 \circ)]_{5}$, then we win the energy.
By Lemma 0, we can win even some more energy by adjusting the distances
between the 2-molecules so that the resulting configuration is
the 2-molecule most homogeneous one:
$[\bullet \bullet (5 \circ) \bullet \bullet (6 \circ)
\bullet \bullet (5 \circ)]_{6}$.

What we have said above, summarized in Lemma 0, 1, and 2,
shows that the strategy that lead to the proof of Theorem 0,
which was concerned with strictly convex interaction energies,
can be applied also to some nonconvex interaction energies
$V_{2}$.\\[2mm]
\noindent
{\bf Theorem 1} ($2$-molecule most homogeneous ground-state
configurations)\\
Consider a lattice gas (\ref{H}) with a nonconvex interaction energy
$V_{2}$ and a particle density $\rho \leq 1/2$.
If $V_{2}(1) = 0$ and $V_{2}(2) \geq 7W/2$, then the ground-state
configurations are the $2$-molecule most homogeneous configurations
of particle density $\rho$.
\vspace{2mm}

\noindent
{\bf Corollary} (Complete devil's staircase)\\
In a lattice gas (\ref{H}) whose interaction energy $V_{2}$ satisfies
the conditions given in Theorem 1,
the particle density versus the chemical potential of particles,
$\rho (\mu)$,
exibits the complete devil's-staircase structure.

\section{ Two lemmas on eliminating atoms and $n$-molecules with
$n\geq3$ }

\noindent
{\bf Lemma 1} (Eliminating of atoms)\\
Consider a lattice gas (\ref{H}) with an interaction energy $V_{2}$,
such that $V_{2}(1) \leq 0$.
Let $X$ be a configuration that does not contain $n$-molecules
with $n \geq 3$.
Suppose that the local configuration $X_{\Lambda}$, where $\Lambda$
is a bounded lattice-connected subset of Z, contains two atoms
separated by $k=0,1,2,\ldots$ molecules.
Then, by rearranging the positions of particles in $X_{\Lambda}$,
it is possible to construct a configuration $Y$ such that
$\rho(Y)=\rho(X)$, $Y_{{\bf Z}- \Lambda}= X_{{\bf Z}- \Lambda}$,
$Y_{\Lambda}$ consists exclusively of $k+1$ $2$-molecules and
$e(Y) < e(X)$.
\vspace{2mm}

\noindent
{\bf Corollary}: Among the configurations that do not contain
$n$-molecules with $n\geq 3$, the lowest energy configuration consists
exclusively of $2$-molecules.
\vspace{2mm}

\noindent
{\bf Proof}:
We propose a proof by reductio ad absurdum. Let $\Lambda$ be a bounded
lattice-connected subset of {\bf Z} and suppose that a local configuration
$X_{\Lambda}$ contains two atoms separated by $k=0,1,2,$ \ldots
$2$-molecules while $X_{{\bf Z} - \Lambda}$ is arbitrary.
The idea is to construct another configuration $Y$,
with $Y_{{\bf Z} - \Lambda}= X_{{\bf Z} - \Lambda}$ and
$Y_{\Lambda}$ consisting of $k+1$ $2$-molecules exclusively, such that
the total energy change, $H(Y,X)$, is negative.

We start with describing $X_{\Lambda}$. Going from left to right,
its $1$-st (or left) atom
is separated by $a$ empty sites from the left environment, after the
first atom we meet the $1$-st $2$-molecule separated from the
$1$-st atom by $n_{1}$ empty sites, later on -- the second $2$-molecule
separated from the $1$-st one by $n_{2}$ empty sites and so on.
The $(i-1)$-th and the $i$-th $2$-molecules are separated by $n_{i}$
empty sites. Finally the $2$-nd (or right) atom is separated from the
last $2$-molecule, i.e., the $k$-th $2$-molecule, by $n_{k+1}$ empty sites
and by $b$ empty sites from the right environment.

As it has been said above, in the new
configuration $Y$ the configurations of the left and right environments
remain unchanged  while $Y_{\Lambda}$ is obtained as follows. We start
with the $1$-st atom and the left particle of the $1$-st $2$-molecule 
in $X_{\Lambda}$ and
put them at the two nearest-neighbor sites located in the center of
the gap of $n_{1}$ empty sites. Clearly, if $n_{1}$ is even, then the
position of the new $2$-molecule (the $1$-st $2$-molecule in
$Y_{\Lambda)}$) is defined uniquely, the center of gravity of the
$2$-molecule coincides with the geometric center of the gap.
On the other hand, if $n_{1}$ is odd, then we can speak of the left
and right central positions of the $2$-molecule. In the first case the
center of gravity of the $2$-molecule is shifted by half the lattice
constant to the left of the geometric center of the gap
while in the second case -- to the right. In the next step we create
the $2$-nd $2$-molecule of $Y_{\Lambda}$ by taking the right particle
of the $1$-st $2$-molecule and the left particle of the $2$-nd
$2$-molecule in $X_{\Lambda}$ and putting them in the center of the gap
made by $n_{2}$ empty sites. The procedure described is continued until
$k+1$ $2$-molecules in $Y_{\Lambda}$ is created. Each time the gap
between the $2$-molecules in $X_{\Lambda}$ is odd we have a choice of
two central positions. Therefore we may end up with as much as $2^{k+1}$
different configurations. In the sequel we shall restrict ourselves to
two simple choices, namely either we always choose the left central position
and the corresponding $Y$ is labeled $Y_{L}$ or we always choose the
right central position what leads to the configuration $Y_{R}$.

Our aim is to estimate the energy changes related to passing from $X$
to $Y_{L}$, $H(Y_{L},X)$ and from $X$ to $Y_{R}$, $H(Y_{R},X)$. 
It is convenient to split the total energy of $X_{\Lambda}$, 
$E_{\Lambda}(X)$, into two parts:
$E_{\Lambda}(X) = E_{\Lambda}^{int}(X) + E_{\Lambda}^{ext}(X)$.
Here $E_{\Lambda}^{int}(X)$ is the {\it internal energy},
that is the sum of pair interaction energies $V_{r_{0}}(r)$ over all pairs
of particles in $X_{\Lambda}$.
The {\it external energy} of $X_{\Lambda}$, $E_{\Lambda}^{ext}(X) $, can
in turn be represented as $E_{\Lambda}^{ext}(X) =
E_{\Lambda}^{L}(X) + E_{\Lambda}^{R}(X)$, where
$E_{\Lambda}^{L}(X)$ is the sum of pair interaction energies over
all pairs of particles consisting of one particle in the left
environment at configuration $X_{{\bf Z} -\Lambda}$ and one particle in
$\Lambda$ at configuration $X_{\Lambda}$, and $E_{\Lambda}^{R}(X)$
is defined similarly.
Therefore
\begin{eqnarray}
\label{ext+int}
H(Y,X) = E_{\Lambda}(Y) - E_{\Lambda}(X) = \nonumber \\
= E_{\Lambda}^{L}(Y) - E_{\Lambda}^{L}(X) +
  E_{\Lambda}^{R}(Y) - E_{\Lambda}^{R}(X) +
  E_{\Lambda}^{int}(Y) - E_{\Lambda}^{int}(X),
\end{eqnarray}
for some $Y$.

The remaining part of our proof consists of two stages. 
In the stage one we study only the external energy differences while 
in the stage two the internal energy differences.\\

\noindent
{\bf Stage one: estimating external energy variation}\\

Consider the pair of particles that constitute the $i$-th $2$-molecule
in $Y_{L}$. Let $E_{i}^{L}(Y_{L})$ and $E_{i}^{L}(X)$ stand for
the external interaction energy of that pair of particles
with the left environment at configurations $Y_{L}$ and $X$,
respectively. Then
\begin{equation}
\label{Lext}
E_{\Lambda}^{L}(Y_{L}) - E_{\Lambda}^{L}(X) =
\sum _{i=1}^{k+1} \left( E_{i}^{L}(Y_{L}) - E_{i}^{L}(X) \right)
\end{equation}
and similarly
\begin{equation}
\label{Rext}
E_{\Lambda}^{R}(Y_{L}) - E_{\Lambda}^{R}(X) =
\sum _{i=1}^{k+1} \left( E_{i}^{R}(Y_{L}) - E_{i}^{R}(X) \right)
\end{equation}
Therefore we are left with estimating the energy differences
$E_{i}^{L}(Y_{L}) - E_{i}^{L}(X)$ and
$E_{i}^{R}(Y_{L}) - E_{i}^{R}(X)$ what requires a detailed
description of positions, in configurations $X$ and $Y_{L}$, 
of the two particles that constitute the
$i$-th $2$-molecule in $Y_{L}$. 
In $X$, the particles mentioned are the right particle of
the $(i-1)$-th $2$-molecule whose position according to the $x$-axis 
is denoted by $x_{i}$ and the left particle of the $i$-th
$2$-molecule whose position is then $x_{i} + n_{i} + 1$. 
Let us recall that the positions mentioned coincide with the distances
between the particles considered and the left environment.
In $Y_{L}$
the corresponding positions are $x_{i} + [n_{i}/2]$ and
$x_{i} + [n_{i}/2] + 1$. 
Now we are ready to express the energies
$E_{i}^{L}(Y_{L})$  and $E_{i}^{L}(X)$ in terms of the interaction
energy, $V_{r_{0}}^{L}(r)$, of a particle at $r \in \Lambda$ with the left
environment (see Section 2 for the definition and properties).
Then
\begin{equation}
\label{L(i)X}
E_{i}^{L}(X) = V^{L}(x_{i}) + V^{L}(x_{i} + n_{i} + 1)
\end{equation}
and
\begin{equation}
\label{L(i)Y_L}
E_{i}^{L}(Y_{L}) = V^{L}(x_{i} + [n_{i}/2]) +
                     V^{L}(x_{i} + [n_{i}/2] + 1).
\end{equation}
Therefore
\begin{eqnarray}
\label{L(i)diff}
E_{i}^{L}(Y_{L}) - E_{i}^{L}(X) = \nonumber \\
\left( V^{L}(x_{i} + [n_{i}/2]) - V^{L}(x_{i})  \right) +
\left( V^{L}(x_{i} + [n_{i}/2] + 1) - V^{L}(x_{i} + n_{i} + 1)
\right).
\end{eqnarray}
Since $V^{L}(r)$ is a decreasing function of distance $r$,
the inspection of the distances that appear in eq.(\ref{L(i)diff})
shows that the
first energy difference is negative while the second one is
positive. Hence the net outcome can only be established by calling
additional properties of $V^{L}(r)$. We use the convexity property of
$V^{L}(r)$ (for $r\geq 2$).
Estimating $E_{i}^{L}(Y_{L}) - E_{i}^{L}(X)$ amounts
to estimating the variations of $V^{L}(r)$ at the intervals
$\left[ x_{i}, x_{i} + [n_{i}/2] \right]$ and 
$\left[ x_{i} + [n_{i}/2] + 1, x_{i} + n_{i} + 1 \right]$ (see sec. 2).
First suppose that $n_{i}$
is even. Then both intervals are of the same length, $n_{i}/2$,
but the second one is shifted to the right with respect to the first one.
Therefore by convexity
\begin{equation}
\label{V_Lconv_ev}
V^{L}(x_{i} + n_{i}/2 + 1) - V^{L}(x_{i} + n_{i} + 1)
\leq
V^{L}(x_{i}) -  V^{L}(x_{i} + n_{i}/2)
\end{equation}
and consequently
\begin{equation}
\label{L(i)diffneg}
E_{i}^{L}(Y_{L}) - E_{i}^{L}(X) \leq 0,
\end{equation}
which is the desired result. However if $n_{i}$ is odd, one of the 
intervals, the one located more to the right, is longer by $1$ and
an analogous reasoning does not reproduce inequality (\ref{L(i)diffneg}).
Translating the longer interval to the left, so that its left end 
coincides with the left end of the shorter one, we obtain by convexity
\begin{equation}
\label{L(i)diffpos}
E_{i}^{L}(Y_{L}) - E_{i}^{L}(X) \leq
V^{L}(x_{i} + [n_{i}/2]) - V^{L}(x_{i} + \{ n_{i}/2 \}).
\end{equation}
{\bf Remark}: In a similar way one can derive a stronger inequality,
namely the r.h.s. of (\ref{L(i)diffpos}) can be replaced by
$V(x_{i} + n_{i}) - V(x_{i} + n_{i} + 1)$.\\

To get an estimate of the external energy
$E^{ext}_{\Lambda}(Y_{L}) - E^{ext}_{\Lambda}(X)$
it remains to find the counterparts of inequalities
(\ref{L(i)diffneg}) and (\ref{L(i)diffpos}) in the case of the right
environment.
For this purpose it is convenient to introduce a second coordinate-axis, 
the $y$-axis, similar to the $x$-axis but pointing in the opposite direction.
We keep using the notions of left and right according to the $x$-axis. 
The zero of the $y$-axis is set at the $1$-st (i.e., most to the left)
particle of the right environment, so that again the $y$-position of a
particle in $\Lambda$ coincides with the distance between that
particle and the right environment.
Now a while of reflection enables us to realize that the interaction energies
with the right environment can be obtained from the interaction energies
with the left environment by replacing $x_{i}$ by $y_{i}$ and interchanging
the square and curly brackets. Thus the counterparts of (\ref{L(i)X}) and
(\ref{L(i)Y_L})  read
\begin{equation}
\label{R(i)X}
E_{i}^{R}(X) = V^{R}(y_{i}) + V^{R}(y_{i} + n_{i} + 1)
\end{equation}
and
\begin{equation}
\label{R(i)Y_L}
E_{i}^{R}(Y_{L}) = V^{R}(y_{i} + \{ n_{i}/2 \}) +
                     V^{R}(y_{i} + \{ n_{i}/2 \} + 1),
\end{equation}
respectively, and the counterpart of (\ref{L(i)diff}) is
\begin{eqnarray}
\label{R(i)diff}
E_{i}^{R}(Y_{L}) - E_{i}^{R}(X) = \nonumber \\
\left( V^{R}(y_{i} + \{ n_{i}/2 \}) - V^{R}(y_{i})  \right) +
\left( V^{R}(y_{i} + \{ n_{i}/2 \} + 1) - V^{R}(y_{i} + n_{i} + 1)
\right).
\end{eqnarray}
Again, if $n_{i}$ is even we find
\begin{equation}
\label{R(i)diffneg}
E_{i}^{R}(Y_{L}) - E_{i}^{R}(X) \leq 0
\end{equation}
and if $n_{i}$ is odd:
\begin{equation}
\label{R(i)diffpos}
E_{i}^{R}(Y_{L}) - E_{i}^{R}(X) \leq
V^{R}(y_{i} + \{ n_{i}/2 \}) - V^{R}(y_{i} + [n_{i}/2]).
\end{equation}
Summing up, the external energy variation on passing from $X$ to $Y_{L}$
has an upper bound of the form
\begin{eqnarray}
\label{TYLext}
E_{\Lambda}^{ext}(Y_{L}) - E_{\Lambda}^{ext}(X) =
E_{\Lambda}^{L}(Y_{L}) - E_{\Lambda}^{L}(X) +
E_{\Lambda}^{R}(Y_{L}) - E_{\Lambda}^{R}(X) = \nonumber \\
\sum _{i=1}^{k+1} \left( E_{i}^{L}(Y_{L}) - E_{i}^{L}(X) \right)
+
\sum _{i=1}^{k+1} \left( E_{i}^{R}(Y_{L}) - E_{i}^{R}(X) \right)
\leq \nonumber \\
\sum _{i=1}^{k+1}
\left( V^{L}(x_{i} + [n_{i}/2]) - V^{L}(x_{i} + \{ n_{i}/2 \}) \right)
+
\left( V^{R}(y_{i} + \{ n_{i}/2 \}) - V^{R}(y_{i} + [n_{i}/2])  \right).
\end{eqnarray}
Again a while of reflection enables us to write down the upper bound for the
change in the total external energy on passing form $X$ to $Y_{R}$, that
represents a counterpart of (\ref{TYLext}). It is enough to interchange
in (\ref{TYLext}) $x_{i}$ with $y_{i}$ and $R$ with $L$:
\begin{eqnarray}
\label{TYRext}
E_{\Lambda}^{ext}(Y_{R}) - E_{\Lambda}^{ext}(X) =
E_{\Lambda}^{L}(Y_{R}) - E_{\Lambda}^{L}(X) +
E_{\Lambda}^{R}(Y_{R}) - E_{\Lambda}^{R}(X) = \nonumber \\
\sum _{i=1}^{k+1} \left( E_{i}^{L}(Y_{R}) - E_{i}^{L}(X) \right)
+
\sum _{i=1}^{k+1} \left( E_{i}^{R}(Y_{R}) - E_{i}^{R}(X) \right)
\leq \nonumber \\
\sum _{i=1}^{k+1}
\left( V^{R}(y_{i} + [n_{i}/2]) - V^{R}(y_{i} + \{ n_{i}/2 \})  \right)
+
\left( V^{L}(x_{i} + \{ n_{i}/2 \}) - V^{L}(x_{i} + [n_{i}/2]) \right).
\end{eqnarray}
By inspection of (\ref{TYLext}) and (\ref{TYRext}) one finds that the upper 
bound given in (\ref{TYLext}) is just the opposite of the upper bound given 
in (\ref{TYRext}). Thus, if 
$E_{\Lambda}^{ext}(Y_{L}) - E_{\Lambda}^{ext}(X) > 0$, then its upper
bound (\ref{TYLext}) is strictly greater than zero, consequently the upper 
bound (\ref{TYRext}) is strictly less than zero what implies that
$E_{\Lambda}^{ext}(Y_{R}) - E_{\Lambda}^{ext}(X) < 0$ and vice versa.
Therefore we proved that 
either $E_{\Lambda}^{ext}(Y_{L}) - E_{\Lambda}^{ext}(X)$
or $E_{\Lambda}^{ext}(Y_{R}) - E_{\Lambda}^{ext}(X)$
is nonpositive.
\vspace{2mm}

\noindent
{\bf Remark}: Let us note here that the lemma is valid also if $k=0$.
\vspace{2mm}

\noindent
{\bf Stage two: estimating internal energy variation}
\vspace{2mm}

Now we are going to estimate the internal energy difference
$E_{\Lambda}^{int}(Y_{L}) - E_{\Lambda}^{int}(X)$, which appears to be
a considerably more laborious task than it was the case for the
external energy difference. To arrive at a satisfactory upper bound
of $E_{\Lambda}^{ext}(Y_{L}) - E_{\Lambda}^{ext}(X)$ we represented
the external energy $E_{\Lambda}^{ext}(Y_{L})$  and
$E_{\Lambda}^{ext}(X)$ by a sum of external energies of pairs of
particles in $\Lambda$ that constitute  $2$-molecules in the configuration
$Y_{L}$ $(Y_{R})$. Such a partition of the external energy led to energy
differences that could be estimated by means of the convexity property of
the interaction involved. We shall need a counterpart of this
partition for $E_{\Lambda}^{int}(Y_{L})$ and $E_{\Lambda}^{int}(X)$.
This time a natural object is not a pair of particles but two pairs of
particles that constitute two $1$-st nearest-neighbor $2$-molecules in
$Y_{L}$ $(Y_{R})$, two $2$-nd nearest-neighbor $2$-molecules, and so on,
until a pair of $(k+1)$-th nearest-neighbor $2$-molecules.
We can define in a natural way the internal interaction energy of those
objects, in terms of $V(r)$. First consider a pair of $1$-st nearest
neighbor $2$-molecules in $Y_{L}$, say the $i$-th and $(i+1)$-th one,
$i=2,\ldots,k-1$ (so that the boundary molecules, the 1-st and the
$(k+1)$-th one, are excluded from our consideration for a moment).
To build up the $i$-th $2$-molecule in $Y_{L}$, we pick up the right
particle of the $(i-1)$-th molecule in $X$ and move it to the right by
distance $[n_{i}/2]$, then the left particle of the $i$-th molecule in
$X$ and move it by distance $\{ n_{i}/2 \}$ to the left. 
Thus the $i$-th and $(i+1)$-th molecules in $Y_{L}$ are separated by
the distance $x^{(1)}_{i} + 1$,
where
\begin{equation}
\label{x1i}
x^{(1)}_{i}:= g(n_{i},n_{i+1}) = \{ n_{i}/2 \} + [n_{i+1}/2],
\end{equation}
and we used the function $g(s,t)$ defined in the Appendix.
The internal interaction energy of the pair of molecules considered
is defined as the sum of all the four interaction energies $V(r)$ 
between the particles separated only by the gap $x^{(1)}_{i}$: 
\begin{equation}
\label{1nni}
E^{(1)}_{i}(Y_{L}) = E_{Y_{L}}(n_{i},n_{i+1}):=
\left( V(x^{(1)}_{i} +1) + V(x^{(1)}_{i} +2) \right) +
\left( V(x^{(1)}_{i} +2) + V(x^{(1)}_{i} +3) \right).
\end{equation}
While according to our description above the definitions
(\ref{x1i}) and (\ref{1nni}) do not apply to these cases
$i=1$ and $i=k$, a direct inspection shows that
\begin{equation}
\label{x11_x1k}
x^{(1)}_{1}= g(n_{1},n_{2}), \;\;\;
x^{(1)}_{k}= g(n_{k},n_{k+1})
\end{equation}
and
\begin{eqnarray}
\label{1nn1_1nnk}
E^{(1)}_{1}(Y_{L}) = E_{Y_{L}}(n_{1},n_{2}):=  \nonumber \\ 
\left( V(x^{(1)}_{1} +1) + V(x^{(1)}_{1} +2) \right) +
\left( V(x^{(1)}_{1} +2) + V(x^{(1)}_{1} +3) \right) \nonumber \\
E^{(1)}_{k}(Y_{L}) = E_{Y_{L}}(n_{k},n_{k+1}):= \nonumber \\
\left( V(x^{(1)}_{k} +1) + V(x^{(1)}_{k} +2) \right) +
\left( V(x^{(k)}_{1} +2) + V(x^{(1)}_{k} +3) \right),
\end{eqnarray}
i.e., these definitions apply to all $i=1,\ldots,k$.
The quantities $x^{(1)}_{i}$ and $E^{(1)}_{i}(Y_{L})$
can be thought of as the values of the functions $g$ and $E_{Y_{L}}$
evaluated at the pairs $(n_{i}, n_{i+1})$, $i=1,\ldots,k$,
of two consecutive gaps in $X_{\Lambda}$, respectively.

A pair of $2$-nd nearest-neighbor molecules in $Y_{L}$, say the
$i$-th and $(i+2)$-th one, is separated by two consecutive gaps
between 1-st nearest-neighbor molecules, $x^{(1)}_{i}$
and $x^{(1)}_{i+1}$.
From the point of view of the interaction energies between the particles 
constituting these molecules, they are separated by one gap of
$x^{(2)}_{i}$ empty sites, where
\begin{equation}
\label{x2}
x^{(2)}_{i}:= x^{(1)}_{i} + x^{(1)}_{i+1} + 2 =
\{ n_{i/2 } \} + [n_{i+2}/2] + (n_{i+1}+2).
\end{equation}
Consequently we can define the internal interaction energy of the
pair considered of $2$-nd nearest-neighbor molecules in $Y_{L}$,
$E^{(2)}_{i}(Y_{L})$,
as the corresponding energy between 1-st nearest neighbors, i.e., by
replacing simply $x^{(1)}_{i}$ in eq.(\ref{1nni}) by $x^{(2)}_{i}$.

While in terms of 1-st nearest-neighbor gaps, $n_{i}$, the pair of 1-st
nearest-neighbor molecules, the $i$-th and $(i+1)$-th one,
was associated with the pair $(n_{i},n_{i+1})$, the pair of 2-nd nearest
neighbor molecules, $i$-th and $(i+2)$-th, is associated with two
consecutive pairs of such pairs
$\left( (n_{i}, n_{i+1}), (n_{i+1}, n_{i+2}) \right)$.
But from the point of view of the interaction energies between the
particles constituting the $i$-th and $(i+2)$-th molecules,
the transition from 1-st nearest-neighbor molecules to 2-nd nearest
neighbor molecules can be described as a ``renormalization process''
that maps two consecutive pairs of consecutive 1-st nearest-neighbor
gaps into a pair of consecutive 2-nd nearest-neighbor gaps:
\begin{equation}
\label{renorm}
\left( (n_{i}, n_{i+1}), (n_{i+1}, n_{i+2}) \right) \longrightarrow
\left( n_{i} + (n_{i+1}+2), n_{i+2} + (n_{i+1}+2) \right).
\end{equation}
Thus the pair of 2-nd nearest-neighbor molecules, the $i$-th
and $(i+2)$-th one,
can also be associated with one pair of gaps -- the pair of 
consecutive 2-nd nearest-neighbor
gaps $\left( n_{i} + (n_{i+1}+2), n_{i+2} + (n_{i+1}+2) \right)$.
Then in terms of the consecutive 2-nd nearest-neighbor gaps
\begin{equation}
\label{renorm_x2}
x^{(2)}_{i} = x \left( n_{i} + (n_{i+1}+2), n_{i+2} + (n_{i+1}+2) \right)
\end{equation}
and
\begin{equation}
\label{2nni}
E^{(2)}_{i}(Y_{L}) =
E_{Y_{L}} \left( n_{i} + (n_{i+1}+2), n_{i+2} + (n_{i+1}+2) \right).
\end{equation}
This construction can be naturally continued:
the pair of 3-rd nearest-neighbor molecules,
the $i$-th and $(i+3)$-th one,
is associated with the pair of consecutive triples
of consecutive 1-st nearest-neighbor gaps
\begin{equation}
\label{3rdnn}
\left(
\left( n_{i},n_{i+1},n_{i+2} \right),
\left( n_{i+1},n_{i+2}, n_{i+3} \right)
\right)
\end{equation}
or, after renormalizing, with the pair of 3-rd nearest-neighbor gaps
\begin{equation}
\label{renorm3rdnn}
\left(
n_{i} + (n_{i+1}+2) + (n_{i+2}+2),n_{i+3} + (n_{i+1}+2) + (n_{i+2}+2)
\right).
\end{equation}
Consequently
\begin{equation}
\label{renorm_x3}
x^{(3)}_{i} = x \left( n_{i} + (n_{i+1}+2)+ (n_{i+2}+2),  
n_{i+2} + (n_{i+1}+2) + (n_{i+2}+2) \right)
\end{equation}
and
\begin{equation}
\label{3nni}
E^{(3)}_{i}(Y_{L}) =
E_{Y_{L}}\left( n_{i} + (n_{i+1}+2) + (n_{i+2}+2),
n_{i+3} + (n_{i+1}+2) + (n_{i+2}+2) \right).
\end{equation}

What we have said above enables us to represent $E_{\Lambda}^{int}(Y_{L})$
as follows:
\begin{equation}
\label{intYL}
E_{\Lambda}^{int}(Y_{L}) = \sum_{i=1}^{i=k} E^{(1)}_{i}(Y_{L}) +
                           \sum_{i=1}^{i=k-1} E^{(2)}_{i}(Y_{L}) +
                           \ldots +
                           E^{(k)}_{1}(Y_{L}),
\end{equation}
where the last component in the above equation stands for the internal
interaction energy of the $1$-st and the last, i.e., $(k+1)$-th molecule
in $Y_{L}$.

Having defined a partition of $E_{\Lambda}^{int}(Y_{L})$ we have to
construct a suitable partition of $E_{\Lambda}^{int}(X)$, so that the
energy differences that will appear eventually in
$E_{\Lambda}^{int}(Y_{L}) - E_{\Lambda}^{int}(X)$ can be shown to be
negative, by means of convexity properties. We have in mind Lemma
A1 of the Appendix.

We start with the level of 1-st nearest neighbors.
Since in $E^{(1)}_{i}(Y_{L})$, given by eq.(\ref{1nni}), the
energies $V(r)$ group naturally into two pairs, so that each distance
involved in one pair has a counterpart differing by 1 in the second pair,
it would certainly be convenient to preserve this property in a partition
of $E_{\Lambda}^{int}(X)$ we are searching for. Then instead of
comparing some four energies in $E_{\Lambda}^{int}(X)$ with the four energies
in $E^{(1)}_{i}(Y_{L})$ we would have to compare only one group of
two energies in $E^{(1)}_{i}(Y_{L})$ with a pair of energies in
$E_{\Lambda}^{int}(X)$.

Let us start with the pair $V(x^{(1)}_{i} +1)$, $V(x^{(1)}_{i} +2)$
of $E^{(1)}_{i}(Y_{L})$.
In order to be able to apply Lemma A1,
the corresponding energies in $E_{\Lambda}^{int}(X)$ should be
associated with some distances, say $r_{1} < r_{2}$, such that
$r_{1} + r_{2} \leq (x^{(1)}_{i} + 1) + (x^{(1)}_{i} + 2)$ and
$r_{1} \leq x^{(1)}_{i} + 1  < x^{(1)}_{i} + 2 \leq r_{2}$.
Searching for such distances is facilitated considerably by Lemma A2,
which gives lower and upper bounds for a quantity like $x^{(1)}_{i}$.
We note that the bounds are the best ones, since they can be attained.
The particles that constitute the $i$-th and $(i+1)$-th $2$-molecules
in $Y_{L}$, whose internal energy is $E^{(1)}_{i}(Y_{L})$,
come from the $i$-th, $(i+1)$-th and $(i+2)$-th $2$-molecules in
$X_{\Lambda}$,
separated by the gaps of $n_{i}$ and $n_{i+1}$ empty sites, or
distances $n_{i}+1$ and $n_{i+1}+1$, respectively.

In order not to consider from the very begining the
specific behaviour of energies at the borders of $X_{\Lambda}$ we
limit the index $i$ to the interval  $2 \leq i \leq k-1$. 
It will be convenient
in the sequel to label the particles that constitute
the $i$-th, $(i+1)$-th and $(i+2)$-th $2$-molecules in $X_{\Lambda}$
by $a,b,c,d,e,f$, in the order from the left to the right.
Thus $r_{a,b}=r_{c,d}=r_{e,f}=1$, where $r_{a,b}$ stands for the
distance between the particles labelled $a,b$, etc.

In what follows we shall label even (odd) integers by the subscript
``$even$'' (``$odd$'').
First suppose that $n_{i}=n_{even}$ and $n_{i+1}=n_{odd}$.
Then by Lemma A4, for $n_{even} < n_{odd}$
\begin{equation}
\label{dist_ineq_eo}
n_{even} + 1 \leq x^{(1)}_{i} + 1 < x^{(1)}_{i} + 2 \leq n_{odd} +1
\end{equation}
and
\begin{equation}
\label{dist_eq_eo}
(n_{even} +1) + ( n_{odd} +1) = (x^{(1)}_{i} + 1) + (x^{(1)}_{i} + 2).
\end{equation}
For $n_{even} > n_{odd}$, $n_{even}$ has to be interchanged with
$n_{odd}$ in ineq.(\ref{dist_ineq_eo}).
Therefore, according to Lemma A4 there is no other choice than
$r_{1}=min\{n_{even} +1,n_{odd}+1 \}$ and
$r_{2}=max\{n_{even}+1,n_{odd} +1 \}$. Fortunately, there are particles
separated by these distances, we can set either $r_{1}=r_{b,c}$ and
$r_{2}=r_{d,e}$ or vice versa. The second group of longer by 1 distances
can be realized as either $r_{1}+1=r_{b,d}$(or $r_{a,c}$)
and $r_{2}+1=r_{c,e}$(or $r_{d,f}$) or vice versa.
Now, by Lemma A1
\begin{equation}
\label{conv_eo}
V(r_{bc}) + V(r_{d,e}) =
V(n_{even}+1) + V(n_{odd}+1) \geq V(x^{(1)}_{i} + 1) + V(x^{(1)}_{i} + 2)
\end{equation}
and a similar inequality is satisfied if $r_{b,c}, r_{d,e}$ are
replaced by $r_{b,d}, r_{c,e}$ (longer by 1) and all other distances in
ineq.(\ref{conv_eo}) are expanded by 1.
Summing up, in the case  $n_{i}=n_{even}$ and $n_{i+1}=n_{odd}$
a good candidate
for the counterpart of $E^{(1)}_{i}(Y_{L})$ in $X_{\Lambda}$ is
\begin{eqnarray}
\label{eni_eo}
E^{(1)}_{i,(e,o)}
& = & \left( V(r_{b,c}) + V(r_{d,e}) \right) +
\left( V(r_{b,d}) + V(r_{c,e}) \right) \nonumber \\
& = & \left( V(n_{even}+1) + V(n_{odd}+1) \right) +
\left( V(n_{even}+2) + V(n_{odd}+2) \right).
\end{eqnarray}
Note that only interactions between 1-st nearest-neighbor
$2$-molecules enter into $E^{(1)}_{i,(e,o)}$, and two one,
out of four possible for each pair of $2$-molecules, have been picked up.

In the second step we suppose that both $n_{i}$ and $n_{i+1}$ are
odd, say $n_{i}=n_{odd}$ and $n_{i+1}=m_{odd}$.
Then by Lemma A4, for $n_{odd} < m_{odd}$
\begin{equation}
\label{dist_ineq_oo}
n_{odd} + 2 \leq x^{(1)}_{i} + 1 < x^{(1)}_{i} + 2 \leq m_{odd}+1,
\end{equation}
while for $n_{odd} < m_{odd}$, $n_{odd}$  and $m_{odd}$ have to be
interchanged in ineq.(\ref{dist_ineq_oo}).
If $n_{odd}= m_{odd}$, then
\begin{equation}
\label{dist_oo}
n_{odd}+1=x^{(1)}_{i} + 1 < x^{(1)}_{i} + 2 = n_{odd}+2.
\end{equation}
Whatever the relation between $n_{odd}$ and $m_{odd}$,
\begin{equation}
\label{dist_eq_oo}
n_{odd} + m_{odd} + 3 = (x^{(1)}_{i} + 1) + (x^{(1)}_{i} + 2).
\end{equation}
Therefore, the distances $r_{1}$ and $r_{2}$ are determined uniquely,
like in the first step.
Trying to realize $r_{1}$ and $r_{2}$ as interparticles distances
we have to be careful not to run into conflict with the choice made
previously. The two consecutive gaps of $n_{odd}$ and $m_{odd}$ empty
sites can be preceded by an even gap, so two interactions between
particles belonging to the pair of 1-st nearest-neighbor
$2$-molecules that are separated by the gap of $n_{odd}$ empty sites
might have already been used.
Therefore, in the case  $n_{i}=n_{odd}$ and $n_{i+1}=m_{odd}$
we choose for the counterpart of $E^{(1)}_{i}(Y_{L})$ in $X_{\Lambda}$
the energy
\begin{eqnarray}
\label{eni_oo}
E^{(1)}_{i,(o,o)}
& = & \left( V(r_{b,d}) + V(r_{d,e}) \right) +
\left( V(r_{a,d}) + V(r_{c,e}) \right) \nonumber \\
& = & \left( V(n_{odd}+2) + V(m_{odd}+1) \right) +
\left( V(n_{odd}+3) + V(m_{odd}+2) \right).
\end{eqnarray}

In the third step we suppose that $n_{i}=n_{odd}$ and
$n_{i+1}=n_{even}$.
Then by Lemma A4, for $n_{odd} < n_{even}$
\begin{equation}
\label{dist_ineq_oe}
n_{odd} + 2 \leq x^{(1)}_{i} + 1 < x^{(1)}_{i} + 2 \leq n_{even} + 2
\end{equation}
and
\begin{equation}
\label{dist_eq_oe}
(n_{odd} +2) + ( n_{even} +2) = (x^{(1)}_{i} + 1) + (x^{(1)}_{i} + 2).
\end{equation}
For $n_{odd} > n_{even}$, $n_{even}$ has to be interchanged with
$n_{odd}$ in ineq.(\ref{dist_ineq_oe}).
Therefore according to Lemma A1 there is no other choice than
$r_{1}=min\{n_{even} +2,n_{odd}+2 \}$ and
$r_{2}=max\{n_{even} +2,n_{odd} +2 \}$.
Therefore in the case  $n_{i}=n_{odd}$ and $n_{i+1}=n_{even}$
we choose for the counterpart of $E^{(1)}_{i}(Y_{L})$ in $X_{\Lambda}$
the energy
\begin{eqnarray}
\label{eni_oe}
E^{(1)}_{i,(o,e)}
& = & \left( V(r_{b,d}) + V(r_{c,e}) \right) +
\left( V(r_{a,d}) + V(r_{c,f}) \right) \nonumber \\
& = & \left( V(n_{odd}+2) + V(n_{even}+2) \right) +
\left( V(n_{odd}+3) + V(n_{even}+3) \right).
\end{eqnarray}
Our definition of $E^{(1)}_{i,(o,e)}$ is not in conflict with those in 
the previous two situations.

In the last step we suppose that both $n_{i}$ and $n_{i+1}$ are even,
say $n_{i}=n_{even}$ and $n_{i+1}=m_{even}$.
Then by Lemma A4, for $n_{even} < m_{even}$
\begin{equation}
\label{dist_ineq_ee}
n_{even} + 2 \leq x^{(1)}_{i} + 1 < x^{(1)}_{i} + 2 \leq m_{even}+1,
\end{equation}
while for $n_{even} > m_{even}$, $n_{even}$  and $m_{even}$ have to be
interchanged in ineq.(\ref{dist_ineq_ee}).
If $n_{even}= m_{even}$, then
\begin{equation}
\label{dist_ee}
n_{even}+1=x^{(1)}_{i} + 1 < x^{(1)}_{i} + 2 = n_{even}+2.
\end{equation}
Whatever the relation between $n_{even}$ and $m_{even}$,
\begin{equation}
\label{dist_eq_ee}
n_{even} + m_{even} + 3 = (x^{(1)}_{i} + 1) + (x^{(1)}_{i} + 2).
\end{equation}
Therefore, again the distances $r_{1}$ and $r_{2}$ are determined uniquely.
As a counterpart of $E^{(1)}_{i}(Y_{L})$ in $X_{\Lambda}$
in the case  $n_{i}=n_{even}$ and $n_{i+1}=m_{even}$
we choose the energy
\begin{eqnarray}
\label{eni_ee}
E^{(1)}_{i,(e,e)}
& = & \left( V(r_{b,c}) + V(r_{c,e}) \right) +
\left( V(r_{b,d}) + V(r_{c,f}) \right) \nonumber \\
& = & \left( V(n_{even}+1) + V(m_{even}+2) \right) +
\left( V(n_{even}+2) + V(m_{even}+3) \right).
\end{eqnarray}
This choice is compatible with the fact that the gap preceding
$n_{even}$ and the one folowing $m_{even}$ can be odd.

So far we have succeeded in selecting a component of
$E_{\Lambda}^{int}(X)$ that is not less than
$\sum_{i=2}^{i=k-1}E^{(1)}_{i}(Y_{L})$.
For any sequence of gaps, $n_{2}$, $n_{3}$, \ldots, $n_{k}$
we have assigned, by means of definitions (\ref{eni_eo}), (\ref{eni_oo}),
(\ref{eni_oe}) and (\ref{eni_ee}),
a sum of pair interactions in such a way that all the four pair
interactions between the particles constituting 1-st nearest-neighbor
2-molecules separated by $n_{i}$, $i=2,3,\ldots,k$, empty sites
enter the sum. So the sum considered can be viewed as the sum over
groups of four two-body interactions between particles of 1-st nearest
neighbor 2-molecules in $X$.

It remains to take care of boundary terms.
Staying at the level of 1-st nearest neighbors,
in $Y_{L}$ they consist
of eight pair interactions: the four ones being the components of
$E^{(1)}_{1}(Y_{L})$
and another four -- the components of $E^{(1)}_{k}(Y_{L})$.
The interactions named can be compared with the following eight components
of $E_{\Lambda}^{int}(X)$: two interactions between the left atom and the
first 2-molecule, two interactions between the 1-st and the 2-nd
2-molecules, two interactions between the $k$-th and $(k+1)$-th
2-molecules and two interactions between the $k$-th 2-molecule and the
right atom. It is sufficient to consider in detail the left end of $\Lambda$.
The interaction of the left atom separated by $n_{1}$ empty sites from
the 1-st molecule amounts obviously to
\begin{equation}
\label{left_at_en}
V(n_{1}+1) + V(n_{1}+2).
\end{equation}
The two still ``free'' interactions between the 1-st and the 2-nd
molecules are:
\begin{eqnarray}
\label{1_2_en}
V(r_{b,c}) + V(r_{a,c}) = V(n_{2}+1) + V(n_{2}+2),
\;\; \mbox{if $n_{2}$ is odd}, \nonumber \\
V(r_{a,c}) + V(r_{a,d}) = V(n_{2}+2) + V(n_{2}+3),
\;\; \mbox{if $n_{2}$ is even}.
\end{eqnarray}
Therefore, we define the left boundary internal energy as:
\begin{eqnarray}
\label{left_bd_en}
E^{(1)}_{1,o} = \left( V(n_{1}+1) + V(n_{2}+1) \right)
+ \left( V(n_{1}+2) + V(n_{2}+2) \right),
\;\; \mbox{if $n_{2}$ is odd}, \nonumber \\
E^{(1)}_{1,e} = \left( V(n_{1}+1) + V(n_{2}+2) \right) +
\left( V(n_{1}+2) + V(n_{2}+3) \right),
\;\; \mbox{if $n_{2}$ is even}.
\end{eqnarray}
Comparing $E^{(1)}_{1,o}$ and $E^{(1)}_{1,e}$ with the energies
$E^{(1)}_{i,(e,o)}$, $E^{(1)}_{i,(o,o)}$, $E^{(1)}_{i,(o,e)}$ 
and $E^{(1)}_{i,(e,e)}$,
defined by eqs.(\ref{eni_eo},\ref{eni_oo},\ref{eni_oe},\ref{eni_ee})
we find that
\begin{equation}
\label{en1_o}
E^{(1)}_{1,o} \left\{
\begin{array}{lll}
=    &  E^{(1)}_{1,(e,o)}, & \mbox{if $n_{1}$ is even} \\
\geq &  E^{(1)}_{1,(o,o)}, & \mbox{if $n_{1}$ is odd}
\end{array}
\right.
\end{equation}
and
\begin{equation}
\label{en1_e}
E^{(1)}_{1,e} \left\{
\begin{array}{lll}
=    &  E^{(1)}_{1,(e,e)}, & \mbox{if $n_{1}$ is even} \\
\geq &  E^{(1)}_{1,(o,e)}, & \mbox{if $n_{1}$ is odd}
\end{array}
\right.
\end{equation}
Thus
\begin{equation}
\label{en1_e_ineq}
E^{(1)}_{1,e} - E^{(1)}_{1}(Y_{L}) \geq 0
\end{equation}
and
\begin{equation}
\label{en1_o_ineq}
E^{(1)}_{1,o} - E^{(1)}_{1}(Y_{L}) \geq 0.
\end{equation}
At the right end the situation is completly analogous. It is enough to 
replace $n_{1}$ by $n_{k}$, $n_{2}$ by $n_{k+1}$ and interchange the 
indices $e$ and $o$ in the formulae (\ref{left_bd_en}, \ref{en1_o}, 
\ref{en1_e}) to arrive at the right boundary internal energies
$E^{(1)}_{k,o}$ and $E^{(1)}_{k,e}$ that are not less than 
$E^{(1)}_{k}(Y_{L})$.

This lengthy sequence of definitions, given above, amounts to the
following: by means of definitions (\ref{eni_eo},\ref{eni_oo},
\ref{eni_oe},\ref{eni_ee},\ref{en1_o},\ref{en1_e}) and the remark
after inequality (\ref{en1_o_ineq}), we defined the function $E_{X}$ on
pairs of consecutive gaps, such that its values on pairs of 1-st
nearest-neighbor gaps, $E^{(1)}_{i}(X):= E_{X}(n_{i},n_{i+1})$,
$i=1,\ldots,k$, satisfy the inequalities
\begin{equation}
\label{en1iX_en1iYL}
E^{(1)}_{i}(X) - E^{(1)}_{i}(Y_{L}) \geq 0.
\end{equation}
Thus $\sum_{i=1}^{k} E^{(1)}_{i}(X)$ can be taken as the required
counterpart of $\sum_{i=1}^{k} E^{(1)}_{i}(Y_{L})$. Now following
the inductive procedure developed to define the partition (\ref{intYL})
of $E_{\Lambda}^{int}(Y_{L})$, we apply $E_{X}$ to pairs of
consecutive 2-nd nearest-neighbor gaps and define
\begin{equation}
\label{en2iX}
E^{(2)}_{i}(X) = E_{X} \left(
n_{i}+ (n_{i+1}+2), n_{i+2}+ (n_{i+1}+2)
\right)
\end{equation}
By Lemma A1 the inequality (\ref{en1iX_en1iYL}) implies that
\begin{equation}
\label{}
E^{(2)}_{i}(X) - E^{(2)}_{i}(Y_{L}) \geq 0,
\end{equation}
so $\sum_{i=1}^{k-1} E^{(2)}_{i}(X)$ can stand for the required
counterpart of $\sum_{i=1}^{k-1} E^{(2)}_{i}(Y_{L})$.
Continuing this process we arrive at the following partition of
$E_{\Lambda}^{int}(X)$:
\begin{equation}
\label{intX}
E_{\Lambda}^{int}(X) = \sum_{i=1}^{k} E^{(1)}_{i}(X) +
\sum_{i=1}^{k-1} E^{(2)}_{i}(X) + \ldots + E^{(k)}_{1}(X) +
V \left( n_{1} + \sum_{i=2}^{k+1}(n_{i}+2) + 1 \right)
\end{equation}
and moreover we obtain that
\begin{equation}
\label{intX_intYL}
E_{\Lambda}^{int}(X) - E_{\Lambda}^{int}(Y_{L}) \geq
V \left( n_{1} + \sum_{i=2}^{k+1}(n_{i}+2) + 1 \right) > 0,
\end{equation}
where $V \left( n_{1} + \sum_{i=2}^{k+1}(n_{i}+2) + 1 \right)$
stands for the interaction of the left atom with the right one.
Clearly, if we substitute $Y_{R}$ for $Y_{L}$ we arrive at the same
conclusion as (\ref{intX_intYL}).
\vspace{2mm}

\noindent
{\bf Remark}: The inequality (\ref{intX_intYL}) holds true if there are
no molecules between two atoms.
\vspace*{5mm}

\noindent
{\bf Lemma 2} (Eliminating of $n$-molecules with $n\geq 3$)\\
Consider a lattice gas (\ref{H}) with an interaction energy $V_{2}$,
such that $V_{2}(1) \geq 0$.
Let $X$ be a configuration containing triples of
particles separated by distance one. Then there is a
configuration $Y$, with $\rho(Y)=\rho(X)$, free of such triples.
Moreover, if $V_{2}(2) \geq (7W-V_{2}(1))/2$,
then $e(Y) < e(X)$, where $W=\sum_{r=3}^{\infty} V_{2}(r)$.
\vspace{2mm}

\noindent
{\bf Corollary}: In a lattice gas of Lemma 2,
if $V_{2}(2) \geq (7W-V_{2}(1))/2$,
then the ground-state configurations do not contain triples of particles
separated by distance one.
In particular, the ground-state configurations
do not contain $n$-molecules with $n \geq 3$.\\

\noindent
{\bf Proof}: The idea of our proof is to start with an arbitrary 
configuration containing forbidden triples of particles, that is
triples of particles separated by distance one, and transform it,
using Lemma 1 and some rearrangements of particles,
whose energy costs are controled by simple bounds,
into another configuration of the same density but free of the forbidden
triples of particles. The proof is completed by showing that the overall
energy cost of attaining the final configuration is negative, provided
that a suitable condition on the interaction energy is imposed.

Let $X$ be an arbitrary configuration with the particle density
$\rho(X) \leq 1/2$ and whose density of the forbidden triples,
$\rho_{3}(X)$, satisfies the inequality $\rho_{3}(X)>0$.
Removing one particle from each
forbidden triple we lower the energy by at least $V_{2}(2) + V_{2}(1)$
and arrive at
a configuration $\tilde{X}$ whose particle density satisfies the double
inequality
\begin{equation}
\label{rhoX_ineq}
1/2 > \rho(\tilde{X}) \geq \rho(X) - \rho_{3}(X).
\end{equation}
Consider the class of configurations that consist of $2$-molecules and
atoms (separated by at least two empty sites) exclusively and whose
density is $\rho(\tilde{X})$.
The configuration $\tilde{X}$ belongs to this class.
Let $Y$ be the lowest energy density configuration in this class,
thus  $e(Y) \leq e(\tilde{X})$.
By (\ref{rhoX_ineq}), $\rho(X) - \rho(\tilde{X}) \leq \rho_{3}(X)$,
therefore the energy density variation $e(X) - e(\tilde{X})$
is bounded from below by
$\left( V_{2}(2) + V_{2}(1) \right) ( \rho(X) - \rho(\tilde{X}) )$.
Consequently, we arrive at the following lower bound for $e(X)$:

\begin{equation}
\label{eX_lb}
e(X) > \left( V_{2}(2)+ V_{2}(1) \right)
( \rho(X) - \rho(\tilde{X}) ) + e(Y).
\end{equation}

By Lemma 1 the configuration $Y$ does not contain atoms,
thus it consists exclusively of $2$-molecules.
It is clear that there is only one (up to translations)
configuration of density $1/2$ that consists exclusively of
$2$-molecules. It is the periodic configuration whose elementary cell 
has the form $[\bullet \bullet \circ \circ]$, where the filled circles
stand, say, for occupied sites. The $2$-molecules are separated
by distance $3$, i.e., by $2$ unoccupied sites. Obviously,
a configuration that consists exclusively of $2$-molecules and whose
particle density is less than $1/2$, contains arrangements of 
$3$ unoccupied sites in a row with a nonvanishing density.
Such a configuration can be thought 
of as a collection of local particle configurations that are separated 
by at least $3$ unoccupied sites in a row.
Shifting the local configurations separated by at least $3$ unoccupied
sites in a row,
while respecting the conditions that they cannot be separated by
less than $2$ unoccupied sites in a row, we can create arrangements of
$6$ unoccupied sites in a row. Then in the middle of the $6$-site gap 
obtained we insert a $2$-molecule. The whole procedure is repeated untill
the particle density comes back to the value $\rho(X)$. 
The result is a configuration  $\tilde{Y}$. 
Now we can try to compare the energy densities $e(\tilde{Y})$ 
and $e(X)$ and see if $e(X)$ can be strictly larger than $e(\tilde{Y})$,
what would prove that ground-state configurations do not contain forbidden
arrangements of particles.
Suppose that we can bound from above by $\Delta$ the energy cost 
of creating a $6$-site gap and inserting there a 2-molecule.
Then we can compare $e(\tilde{Y})$ and $e(Y)$:
\begin{equation}
e(\tilde{Y}) - e(Y) \leq \Delta ( \rho(X) - \rho(\tilde{X}) )/2.
\end{equation}
By means of inequalities (\ref{rhoX_ineq}) and (\ref{eX_lb}), we arrive
at the folowing inequality relating $e(X)$ and $e(\tilde{Y})$:
\begin{equation}
e(X) > e(\tilde{Y}) + (V_{2}(2) + V_{2}(1) - \Delta /2)
( \rho(X) - \rho(\tilde{X}) ).
\end{equation}
Therefore, the ground-state configurations do not contain forbidden
arrangements of particles if $V_{2}(2)+V_{2}(1) \geq \Delta /2 $.\\

It remains to derive the upper bound $\Delta$.
For that we first estimate the energy cost of translating a single particle
by distance $1$. Let a chosen particle be separated from its left nearest
neighbor by distance $r$ and from its right nearest neighbor by
distance $r^{'}$.
The total interaction energy of the particle can be viewed as the sum of
the interaction energies
$E_{L}^{(1)}(r)$ and $E_{R}^{(1)}(r^{'})$ with
all the particles to its left and to its right, respectively.
Suppose we shift our particle by distance $1$ to the left.
Then the change in the total energy of the particle is bounded from
above by $E_{L}^{(1)}(r - 1) - E_{L}^{(1)}(r)$, since
$E_{R}^{(1)}(r^{'} + 1) - E_{R}^{(1)}(r^{'}) \leq 0$.
Let $x_{i}$, $i=1,2,\ldots $, be the distance between the $i$-th left
nearest neighbor and the $(i + 1)$-th left nearest neighbor. Then
\begin{eqnarray}
\label{left_en_ineq}
E_{L}^{(1)}(r - 1) - E_{L}^{(1)}(r) = 
\left(
V_{2}(r-1) + V_{2}(r-1+x_{1}) + V_{2}(r-1+x_{1}+x_{2}) + \ldots \right)
\nonumber \\
- \left(
V_{2}(r) + V_{2}( r+x_{1}) + V_{2}(r+x_{1}+x_{2}) + \ldots \right)
= V_{2}(r-1)
+ \nonumber \\
 - \left( V_{2}(r) - V_{2}(r-1+x_{1}) \right)
- \left( V_{2}( r+x_{1}) - V_{2}(r-1+x_{1}+x_{2}) \right) - \ldots
\leq V_{2}(r-1).
\end{eqnarray}
By means of inequality (\ref{left_en_ineq}) we are able to give a
simple upper bound
for the energy cost of translating a local configuration considered
above by distance $1$ to the left (right). This energy does not exceed
the sum of $V_{2}(r-1)$ over all distances $r$, $r \geq 4$,
between the particles of the configuration
considered and the left nearest neigbor particle of this
configuration:
\begin{equation}
\sum_{r=4}^{\infty} n_{r} V_{2}(r-1) \leq \sum_{r=4}^{\infty} V_{2}(r-1) = W,
\end{equation}
where $n_{r}=1$ if the local configuration considered contains a
particle separated by distance $r$ from the left nearest-neighbor
particle of this configuration, and $n_{r}=0$ otherwise.
Consider again a configuration that is a collection of local
configurations separated by at least $3$ unoccupied sites.
We can create a $6$-site gap between two local configurations
by shifting a chosen local configuration
by distance one to the left, then shifting its right nearest-neighbor
local configuration by distance 1 to the right,
what creates a $5$-site gap between the configuration chosen and its
right nearest-neighbor local configuration,
and after that shifting simultaneously the right nearest-neighbor
and the right next nearest-neighbor local configurations
by distance 1 to the right.
Each of the three reshuffles of local configurations increases
the energy by no more than $W$.
Thus, a $6$-site gap can be created at the cost that does not exceed
$3W$. Now inserting a $2$-molecule in the middle of the $6$-site gap
created will cost no more than $4W + V_{2}(1)$.
Therefore, we can set $\Delta=7W + V_{2}(1)$.

\section{Some extensions of the main result and applications}

The both lemmas, formulated and proved in the previous section,
set conditions on the value of $V_{2}$ at distance one, which go
in opposite directions. It is for this reason that in Theorem 1
this value is set to zero.
However, it is not hard to see that with some effort this condition
can be relaxed.

First, suppose $V_{2}(1) < 0$. This condition fits the
assumptions of Lemma 1 but not those of Lemma 2.
In the latter case, the energy gain on breaking forbidden triples
of particles is as before $V_{2}(2)+V_{2}(1)$
but is diminished now by the second term and may even become negative.
On the other hand, the energy increase on inserting 
a 2-molecule into a six-site gap, 
$4W + V_{2}(1)$, is also diminished.
Thus, Theorem 1 holds true also for $V_{2}(1) < 0$, but such that
$V_{2}(2) \geq (7W- V_{2}(1))/2$, which can only take place if
$|V_{2}(1)| < 2V_{2}(2)$.

Second, suppose $V_{2}(1) > 0$. Now, in turn, this condition fits  the
assumptions of Lemma 2 but not those of Lemma 1, which is a more
severe problem to overcome than the previous one. The final conclusion
of our proof of Lemma 1 is a consequence of the fact that the energy
difference of an initial local configuration,
consisting of two atoms separated by some $k$ 2-molecules,
and the final one, consisting of $k+1$ 2-molecules, is not smaller
than $V_{2}(r_{atom}) - V_{2}(1)$, where $r_{atom}$ is an arbitrary
distance between the two atoms in the initial local configuration.
If, as we have supposed originally, $V_{2}(1)=0$, then the lower bound
obtained implies that the energy difference considered is positive
what completes our proof of Lemma 1. If however $V_{2}(1) > 0$,
to arrive at the same conclusion by means of a similar proof
we would have to limit from above $r_{atom}$, which is not possible.
Instead, we can turn to a strategy analogous to that used in Lemma 2.

In order to restrict the set of possible ground-state configurations
to the set $C^{1}_{2,\rho}$ we have to get rid of forbidden pairs
of particles, i.e., particles separated by distance $1$. We fix 
distance $r\geq 4$ and choose a particle density $\rho < 1/r$.
Removing one
particle from each forbidden pair we lower the energy by $V_{1}(1)$
and arrive at a configuration of lower density, consisting of atoms
exclusively. The ground-state configuration among such configurations
is a most homogeneous configuration, in which the smallest distance $d$
between neighboring atoms is greater than $r$. Then we insert back the
removed particles in such a way that the smallest distance between
neighboring atoms is greater than $[d/2]+1$. Now, if
$V_{1}(1) > 2 \sum_{r\geq [d/2]+1} V_{1}(r)$, then the obtained configuration
has a lower energy density. Therefore, under the stated conditions, 
any ground-state configuration consists only of atoms and
consequently is a most homogeneous configuration.

Finally, we would like to point out that our Theorem 1 and its
generalizations discussed above can be used to study properties of
two- and three-dimensional lattice gases.

Consider first a two-dimensional
lattice gas on the square lattice ${\bf Z}^{2}$. Let $(x,y)$ be an
orthogonal coordinate system, such that the its axes coincide with
lattice directions with unit lattice constants. Suppose that with
respect to this coordinate system the Hamiltonian $H_{\Lambda}$
decomposes into $H^{x}_{\Lambda} + H^{y}_{\Lambda}$, where
$H^{x}_{\Lambda}$ is a sum of two-body potentials supported by pairs
of sites along the $x$-direction, and $H^{y}_{\Lambda}$ is an
analogous sum but in the $y$-direction.
Then, knowing the
one-dimensional ground-state configurations of $H^{x}_{\Lambda}$,
we can easily construct two-dimensional ground-state configurations
of $H_{\Lambda}$. For instance, the particle configurations of lattice
lines in the $x$-direction can be chosen as the
ground-state configurations of $H^{x}_{\Lambda}$, with the condition
that, going in the positive $y$-direction, the configuration of
the next lattice line is shifted by one lattice constant in the
positive $x$-direction. The Hamiltonians considered above naturally
appear in surface physics \cite{ishimura}.

Another situation, where our results are applicable, is the realm of
three-dimensional layered systems, like those studied extensively
by Fisher and collaborators, see for instance \cite{fisher}.
\vspace{2mm}

\noindent {\bf Acknowledgments}. We would like to thank the Polish
Committee for Scientific Research, for a financial support under
the grant KBN 2P03A01511. One of the authors (J.\ J.) is grateful to
Ch.\ Gruber for discussions of the obtained results and kind
hospitality during his stay in the Institut de Physique Th\'{e}orique
of the Ecole Polytechnique F\'{e}d\'{e}rale de Lausanne.

\section{Appendix}

{\bf Definition 1} We say that a real function $f$, defined on integers
$n \geq n_{0}$, is {\it convex}, if for every $n \geq n_{0} + 1$ the
following inequality is satisfied:
\begin{equation}
\label{conv_def}
f(n+1) + f(n-1) \geq 2 f(n).
\end{equation}
\noindent
{\bf Lemma A1} (Convexity of decreasing $f$ in {\bf Z})\\
Let integers $x,y,s,t$ satisfy the inequalities
\begin{equation}
\label{points}
n_{0} \leq x \leq s < t \leq y  \; \; and \; \; x+y \leq s+t
\end{equation}
and let $f$ be a convex and decreasing function for integers
$n \geq n_{o}$. Then
\begin{equation}
\label{conv_prop}
f(x) + f(y) \geq f(s) + f(t).
\end{equation}
\noindent
{\bf Definition 2} For any positive integer $n$, $[n/2]$ is the greatest
integer that does not exceed $n/2$ while $\{ n/2 \} = n - [n/2]$.\\

\noindent
{\bf Lemma A2} (Properties of bracket functions $[.]$ and $\{.\}$)\\
For positive integers $n \leq m$,
$ [\frac{n}{2}] \leq [\frac{m}{2}] $, and
$ \{\frac{n}{2}\} \leq \{\frac{m}{2}\} $. 

Moreover,
\begin{equation}
\label{additivity_sq_br}
\left[ \frac{m+n}{2} \right]= 
\left\{
\begin{array}{ll}
[\frac{m}{2}] + [\frac{n}{2}] + 1, &  \mbox{if $m$ and $n$ are odd} \cr
[\frac{m}{2}] + [\frac{n}{2}],     &  \mbox{otherwise}
\end{array}
\right.
\end{equation}
and
\begin{equation}
\label{additivity_curl_br}
	\left\{ \frac{m+n}{2} \right\}= 
\left\{
\begin{array}{ll}
\{\frac{m}{2}\} + \{\frac{n}{2}\} - 1, & \mbox{if $m$ and $n$ are odd} \\
\{\frac{m}{2}\} + \{\frac{n}{2}\},     & \mbox{otherwise.}
\end{array}
\right.
\end{equation}

\noindent
{\bf Definition 3} For an ordered pair of positive integers $(s,t)$
we define 
\begin{equation}
\label{g_def}
g(s,t) = \{ s/2 \} + [t/2].
\end{equation}

\noindent
{\bf Lemma A3} (Properties of function $g$)\\
For positive integers $s \leq t$,
\begin{equation}
\label{g_bounds}
s \leq g(s,t) \leq t.
\end{equation}
More refined bounds are given in Lemma A4.
For positive integers $p,s,t$
\begin{equation}
\label{g_shift}
g(s+p,t+p) = g(s,t) + p.
\end{equation}

\noindent
{\bf Lemma A4} (Best bounds for function $g$)\\
Let $n_{even}$ and $m_{even}$ be two positive even integers and
$n_{odd}$ and $m_{odd}$ -- two positive and odd integers and let
$g_{e,o}$ be an abreviation for $g(n_{even}, n_{odd})$, etc.
Then \\[2mm]
$
\begin{array}{lrrcllrcccr}
EO: (n_{even}, n_{odd}) & & & & & & & & & & \\
(EO1) & If & n_{even} & < & n_{odd}, & then  &
n_{even} & \leq & g_{e,o} & \leq & n_{odd} -1 \\
(EO2) & If & n_{even} & > & n_{odd}, & then  &
n_{odd} & \leq & g_{e,o} & \leq & n_{even} -1
\end{array}
$
\begin{equation}
2g_{e,o} = n_{even} + n_{odd} - 1
\end{equation}
$
\begin{array}{lrrcllrcccr}
OO: (n_{odd}, m_{odd}) & & & & & & & & & & \\
(OO1) & If & n_{odd} & < & m_{odd}, & then  &
n_{odd} + 1 & \leq & g_{o,o} & \leq & m_{odd} -1 \\
(OO2) & If & n_{odd} & > & m_{odd}, & then &
m_{odd} + 1 & \leq & g_{o,o} & \leq & n_{odd} -1  \\
(OO3) & If & n_{odd} & = & m_{odd}, & then  &
            &       & g_{o,o} &  =  & n_{odd}=m_{odd}
\end{array}
$
\begin{equation}
2g_{o,o} = m_{odd} + n_{odd}
\end{equation}
$
\begin{array}{lrrcllrcccr}
OE: (n_{odd}, n_{even} ) & & & & & & & & & & \\
(OE1) & If & n_{odd} & < & n_{even}, & then &
n_{odd} + 1 & \leq & g_{o,e} & \leq & n_{even} \\
(OE2) & If & n_{odd} & > & n_{even}, & then &
n_{even} + 1 & \leq & g_{o,e} & \leq & n_{odd} \\
\end{array}
$
\begin{equation}
2g_{o,e} = n_{odd} + n_{even} + 1
\end{equation}
$
\begin{array}{lrrcllrcccr}
EE: (n_{even}, m_{even}) & & & & & & & & & & \\
(EE1) & If & n_{even} & < & m_{even}, & then &
n_{even} + 1 & \leq & g_{e,e} & \leq & m_{even} -1 \\
(EE2) & If & n_{even} & > & m_{even}, & then &
m_{even} + 1 & \leq & g_{e,e} & \leq & n_{even} -1
\end{array}
$
\begin{equation}
2g_{e,e} = m_{even} + n_{even}
\end{equation}

\end{document}